\documentclass[prl,aps,twocolumn]{revtex4-1}

\usepackage{CJK}
\usepackage{mathrsfs}
\usepackage{amssymb}
\usepackage{graphicx}
\usepackage{bm}
\usepackage{graphicx}
\usepackage{float} 
\usepackage{longtable}
\usepackage{amsmath}
\usepackage{color}
\usepackage{multirow}
\usepackage{rotating}
\usepackage{upgreek}
\newfont{\largemi}{cmmi10}
\newfont{\smallmi}{cmmi6}

\def\eqref#1{Eq.~(\ref{eq:#1})}

\makeatletter

\begin{document}
\setlength{\abovedisplayskip}{1ex} 
\setlength{\belowdisplayskip}{1ex} 

\title{Unmixing symmetries}

\author{Calvin W. Johnson}
\affiliation{Department of Physics, San Diego State University, 5500 Campanile Drive, San Diego, CA 02182-1233, United States}
\date{\today}
\begin{abstract}
 The low-lying spectra of atomic nuclei display diverse behaviors, for example rotational bands, which can be described 
 phenomenologically by simple symmetry groups such as spatial SU(3). This leads to the idea of \textit{dynamical symmetry}, 
 where the Hamiltonian commutes with the  Casimir operator(s) of a group, and is  block-diagonal in subspaces defined by the group's irreducible representations
 or irreps.
Detailed microscopic calculations, however, show these symmetries are in fact often strongly mixed and the wave function fragmented across many irreps. More commonly 
the fragmentation across members of a band are similar, which is called a
 \textit{quasi-dynamical symmetry}.  In this Letter I explicitly, albeit numerically, construct unitary transformations from a quasi-dynamical symmetry
to a dynamical symmetry, adapting the \textit{similarity renormalization group,} or SRG, in order to transform away the symmetry-mixing parts of the Hamiltonian.  The standard SRG produces unsatisfactory results, 
forcing the induced dynamical symmetry to be dominated by high-weight irreps irrespective of the original decomposition. Using \textit{spectral distribution theory} to 
rederive and diagnose standard SRG, I introduce a new form of SRG. The new SRG transforms a quasi-dynamical symmetry to a dynamical 
symmetry, that is, unmixes the mixed symmetries, with intuitively more appealing results.
\end{abstract}
\maketitle

The spectra of atomic nuclei display a rich portfolio of behaviors, the most striking of which are rotational and vibrational bands. These can  be elegantly described using 
spectrum-generating algebras whose eigenspectra as well as transition probabilities (up to an overall scale), capture experimental data. 
This leads to the idea of a \textit{dynamical symmetry} \cite{talmi1993simple,rowe2010fundamentals}, 
marked by the Hamiltonian commuting with the group's Casimir operator and the wave functions  wholly contained within a single irreducible representation (irrep) of 
the underlying group.
Dynamical symmetries of this kind are mostly invoked in nuclear structure physics, with some discussions in atomic and molecular physics \cite{wulfman1973approximated,PhysRevA.18.1770,iachello1995algebraic,cederbaum1995symmetry,PhysRevLett.116.123001}.



The problem is, microscopic calculations showing true dynamical symmetries are rare. Standard pieces of the nuclear force, such as spin-orbit 
splitting and pairing \cite{rochford1988survival,Bahri1995171,escher1998pairing,PhysRevC.63.014318} strongly break the symmetry and mix irreps.  This is puzzling in light of the fact that one can \textit{empirically} use algebraic methods to reproduce data.
A further piece of the puzzle is the existence of \textit{quasi-dynamical symmetries} \cite{PhysRevC.58.1539,rowe2000quasi,bahri20003}, where the pattern of mixing symmetries, although often very complex, is similar across members of a 
band. 

In this Letter I  adapt a method, the similarity renormalization group (SRG), to generate a unitary transformation that largely \textit{unmixes} the symmetry.  (I use `mixing' rather than 'breaking' symmetry because the former better matches the continuous process described below.)    The standard SRG, however, 
produces for some states unsatisfactory results, so I introduce a novel variant of SRG which provides more intuitively appealing results.  Thus  I can  transform away the symmetry-mixing 
terms in  a Hamiltonian. As a bonus, a new light is shed on  the behavior of SRG, a widely used method.

\begin{figure}
\centering
\includegraphics[width=0.4\textwidth,clip]{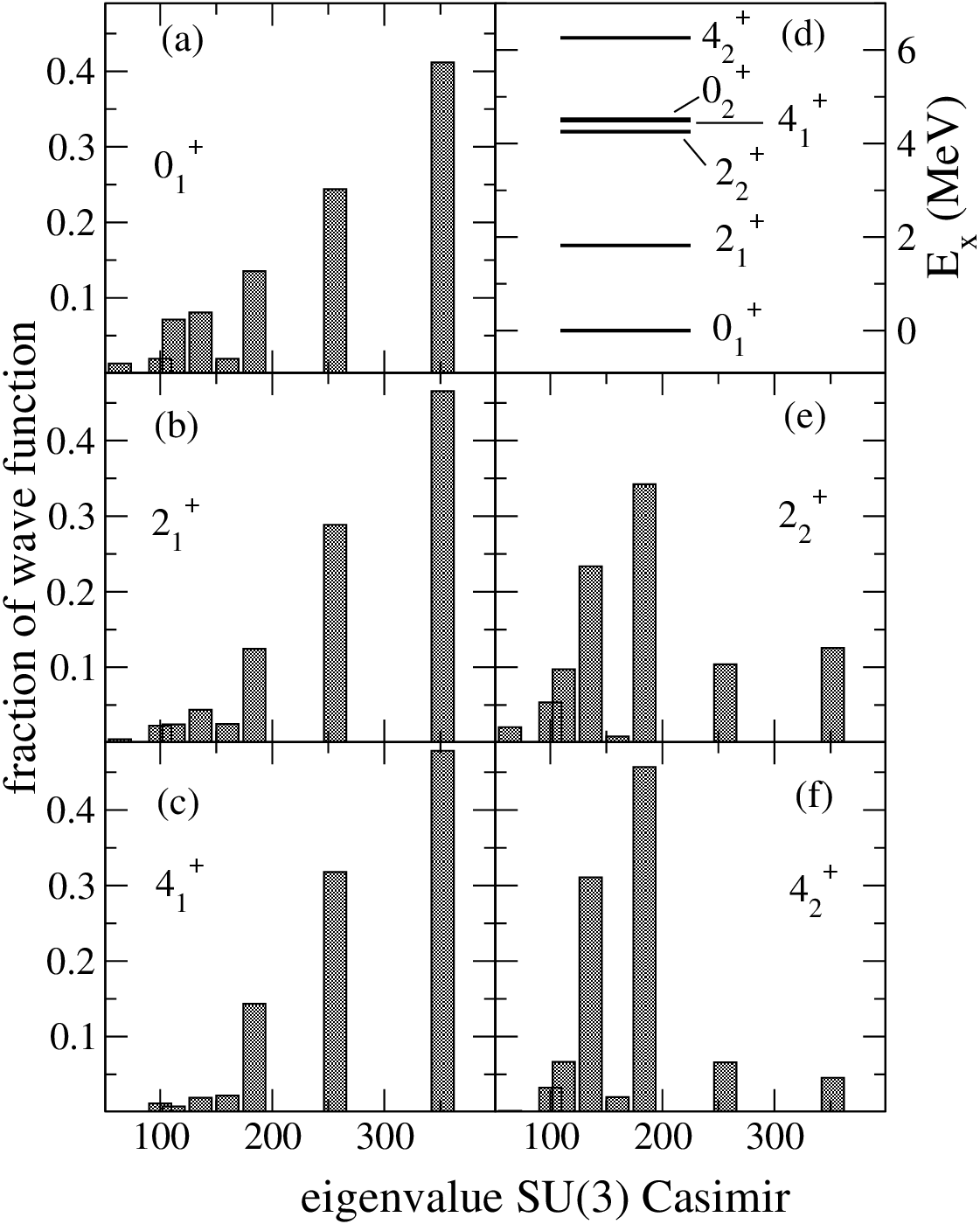}
\caption{ 
Low-lying levels and SU(3) decompositions for $^{36}$Ar in the $sd$ shell using 
the USDB interaction. Panels (a)-(c) show the ground state band, while panels (e)-(f) 
the excited $\gamma$-band.  Panel (d) shows the calculated excitation energies.}
\label{ar36base}
\end{figure}

To illustrate the mixing of symmetries I   decompose
nuclear wave functions, calculated via configuration-interaction, into subspaces defined by irreducible representations.  Let $\hat{C}$ be a Casimir operator for a group, and let $z$ denoted eigenvalues of the Casimir, so that 
$\hat{C}|z, \alpha \rangle = z | z, \alpha \rangle$. The eigenvalues are  highly degenerate and can label subspaces
or irreducible representations. An familiar example is the rotation group, with the Casimir $\hat{J}^2$ with eigenvalues $j(j+1)$ labeling
subspaces of good total angular momentum. For a given state $| \Psi \rangle$, define the fraction of the wave function in a subspace 
labeled by $z$ as 
\begin{equation}
f(z) = \sum_{\alpha} \left | \langle z, \alpha | \Psi \rangle \right |^2
\end{equation}
For dynamical symmetries, $f(z)=1$ for some value of $z$, and zero for all other values. For any state, however, one can calculate 
$f(z)$ efficiently 
\cite{PhysRevC.63.014318,PhysRevC.91.034313,PhysRevC.95.024303}.

Fig.~\ref{ar36base} shows calculations of $^{36}$Ar in  
 the   $1s_{1/2}$- $0d_{3/2}$-$0d_{5/2}$ or $sd$ shell, 
which has a frozen $^{16}$O core, 
using the phenomenological universal $sd$ interaction version 
`B' (USDB)~\cite{PhysRevC.74.034315}, which I decomposed using the quadratic SU(3) Casimir, 
$\hat{C}_2 = \vec{Q} \cdot \vec{Q} + 3 L^2 $, where $L$ is orbital angular momentum 
and $\vec{Q}$ is the so-called Elliott quadrupole operator. The eigenvalues of $\hat{C}_2$ can be expressed in terms of integer quantum numbers $\lambda$ and $\mu$, 
$4(\lambda^2 + \lambda\mu + \mu^2 + 3\lambda + 3\mu)$ \cite{talmi1993simple}.
Because I use only one of two SU(3) Casimirs, 
the decompositions are in many cases  sums of irreps.  One can interpret the results in terms of $( \lambda, \mu)$  of SU(3), 
but I leave those off precisely because those details, while of interest to the specialist, are irrelevant to the points being made here. 
 I chose $^{36}$Ar because it is 
tractable for the following calculations, has strong mixing yet  clearly demonstrates a quasi-dynamical symmetry.  Other nuclides 
show similar results.

Note that the pattern of  fragmentation of the wave function over irreps 
is  repeated across several states \cite{rochford1988survival,PhysRevC.63.014318}. 
This is an example of {quasi-dynamical symmetry,} 
which turns out to be surprisingly 
commonplace.

Seeing the repeated patterns of quasi-dynamical symmetries, it is natural to wonder if one could transform away the 
symmetry-mixing terms to regain a true dynamical symmetry.  While it's not yet known how to choose analytically such a 
unitary transformation,  there does exist a well-known method for numerically constructing unitary transformations, 
the \textit{similarity renormalization group,} or SRG  
\cite{PhysRevD.48.5863,wegner1994flow,PhysRevC.75.061001,PhysRevLett.103.082501,bogner2010low,tsukiyama2011medium,hergert2016medium}.
SRG is widely used in nuclear physics to soften nuclear forces  for \textit{ab initio} calculations, by approximately decoupling low-momentum and high momentum 
states, with analogous  applications in atomic and molecular physics \cite{evangelista2014driven}, non-relativistic reduction of the Dirac 
equation \cite{PhysRevC.85.021302}, and particle physics \cite{PhysRevD.67.045001}.
In each case one uses SRG to approximately decouple a model space from the rest of the space to improve convergence of calculations.
Here I present a novel use of SRG to decouple or unmix group symmetries.  
The beauty of this approach is that it does not require explicit knowledge of the origin of symmetry mixing.

Consider a parameterized unitary transformation of a Hamiltonian, $\hat{H}(s) = \hat{U}(s) \hat{H} \hat{U}^\dagger(s)$, and 
let $\hat{\eta} = (d \hat{U}(s)/ds)  \hat{U}^\dagger(s)$ be an anti-Hermitian operator. Then one can construct an equation for unitary evolution,
\begin{equation}
\frac{d \hat{H}(s)}{ds} = \left [ \hat{\eta}, \hat{H}(s) \right ]. \label{genericU}
\end{equation}
For standard SRG, one introduces a fixed Hermitian operator called the \textit{generator}, $\hat{G}$, and then choose 
\begin{equation}
\hat{\eta} = \left [ \hat{G}, \hat{H}(s) \right ].
\label{stdSRG}
\end{equation}

\begin{figure}
\centering
\includegraphics[width=0.4\textwidth,clip]{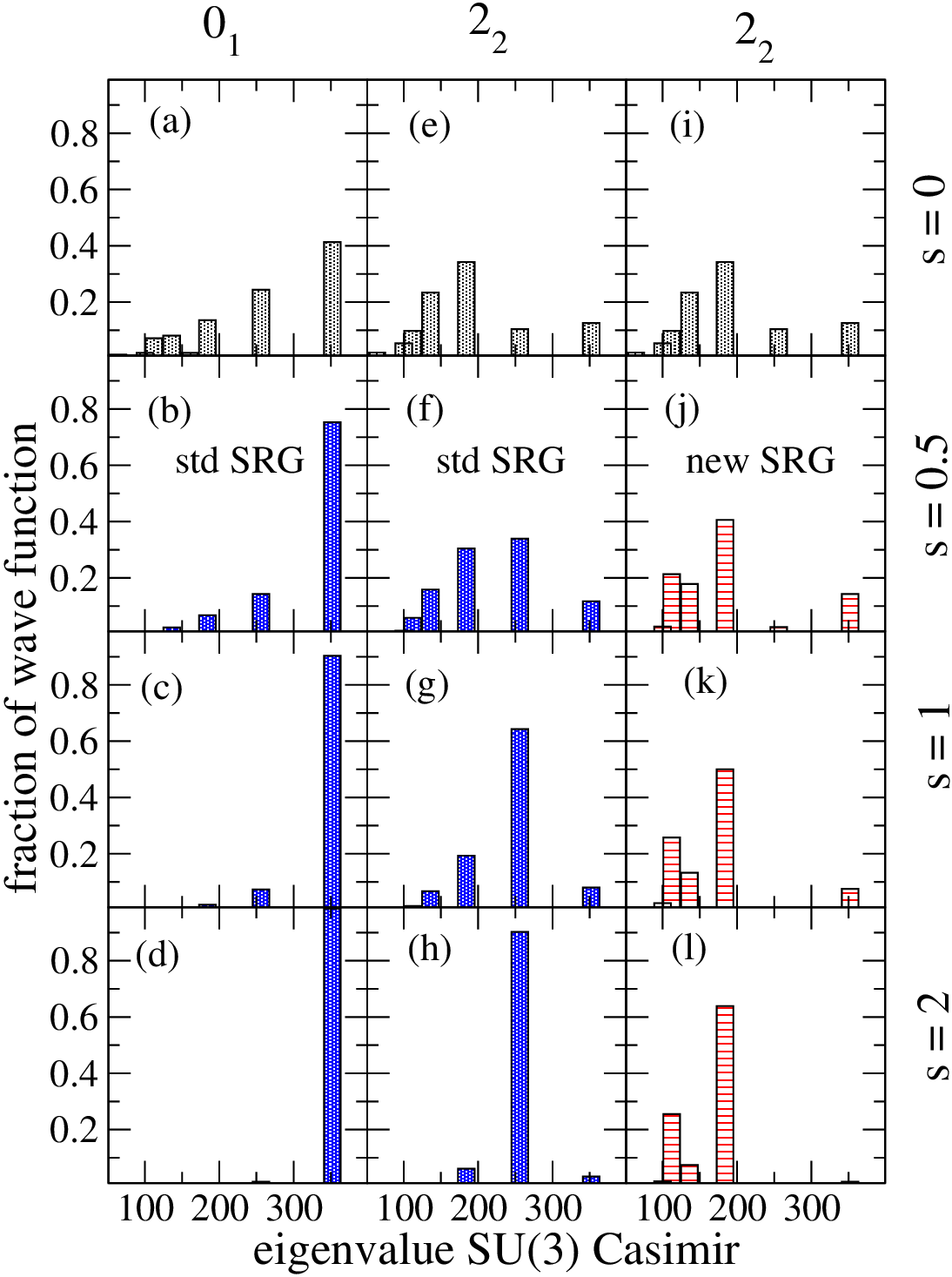}
\caption{ 
Decomposition of the $0_1^+$ and $2_2^+$ states of $^{36}$Ar, using original ($s=0$) and SRG-evolved ($s> 0 $) Hamiltonians. Here $s$ is dimensionless.
The top row, panels (a), (e) and (i)), gives the decompositions for states from unevolved Hamiltonians. The left column, panels (a)-(d), 
shows the evolution under the standard SRG for the $0_1^+$ state.  The middle column, panels (e)-(h), shows the evolution under 
standard SRG for the $2^+_2$ states, while the right column, panels (i)-(l), shows the evolution under the new SRG for the $2^+_2$ state.
Not shown is the evolution of the $0_1^+$ state under the new SRG, which is nearly indistinguishable to that under standard SRG.
\label{ar36srg} }
\end{figure}

To soften the nuclear interaction, one typically uses the kinetic energy operator $\hat{T}$ as the generator; there are other generators 
for other applications, such as the in-medium SRG \cite{tsukiyama2011medium,hergert2016medium}.  Instead here I chose $\hat{G}$ to be $-\hat{C}_2$ of SU(3) (the minus sign is 
because one knows \cite{ring2004nuclear} that $- \vec{Q}\cdot \vec{Q}$ is an approximate component of the nuclear force), although 
in principle one could use any group Casimir.  In order to 
ensure exact unitarity, I act directly on the many-body matrix, which here is of dimension 640; thus the energy spectrum is unchanged, which was confirmed after evolution.   
The differential equation is solved using 
fourth-order Runge-Kutta. (There are more sophisticated methods for solving SRG \cite{PhysRevC.92.034331}, but Runge-Kutta is straightforward to code.)  Because the SU(3) Casimir has no meaningful dimensions, I  rescaled $\hat{\eta}$ so that the two-norm $|| \hat{\eta} || = 1$, and 
the evolution parameter $s$ is dimensionless. 

Fig.~\ref{ar36srg} shows decompositions for two states, the $0_1^+$ ground state and the $2_2^+$ state, as the Hamiltonian is evolved under SRG, starting at $s=0$ along the 
top row, and then increasing to $s=2$ along the bottom row.  While the Hamiltonian is evolved, the decomposition was performed using the original SU(3) Casimir.
The left column shows the evolution for the $0_1^+$ ground state under the ``standard'' SRG, which uses 
Eq.~(\ref{stdSRG}), while the middle column shows the same for the $2_2^+$ state.  In both  cases the decomposition evolves to a single irrep, that is, dynamical symmetry. 

Yet upon closer inspection, something goes `wrong'  under SRG evolution. While the ground state essentially has all its strength going into 
the irrep which already has the largest fraction, as one might expect or at least hope for,  the $2^+_2$ state goes to a higher-weight irrep barely occupied in the 
original decomposition.
?

Why does SRG drive the fractional distribution to the ``wrong'' irreps? To understand this, I borrow concepts from \textit{spectral 
distribution theory} or SDT  \cite{french1967measures,french1983statistical,wong1986nuclear}.  A key idea in SDT is the introduction of an inner product on a linear space of  Hermitian operators, 
represented by finite Hermitian matrices with  dimension $N$. For two such operators, $\mathbf{A}, \mathbf{B}$ ( here on I use boldface type to emphasize they are finite matrices), the inner product is
\begin{equation}
(A,B) = \frac{1}{N} \mathrm{tr} \mathbf{A \,B} - \frac{1}{N} \mathrm{tr} \mathbf{A}\, \frac{1}{N} \mathrm{tr} \mathbf{B}.
\label{innerproduct}
\end{equation}
With an inner product one can define how close or different two Hermitian operators are, and even define an ``angle'' between 
two interactions.

Now suppose we want a unitary transformation on a Hamiltonian $\mathbf{H}$ that makes it as close as possible to the generator $\mathbf{G}$.
Because $N$ is fixed, $\mathbf{G}$ is fixed, and, by unitarity, $\mathrm{tr} \, \mathbf{H}$ is fixed, this means we want to maximize 
$\mathrm{tr} \, \mathbf{G\, H}(s)$.  While guaranteeing a global maximum is not trivial, let us suppose we follow the generic evolution equation 
(\ref{genericU}) and choose to maximize the \textit{rate} at which 
the unitary transformation increases $\mathrm{tr}\, \mathbf{G\,H}(s)$, that is, we want to maximize
\begin{equation}
\frac{d}{ds} \mathrm{tr} \,\mathbf{G\,H}(s) =
\mathrm{tr} \,\mathbf{G} \frac{d}{ds} \mathbf{ H}(s) =
 \mathrm{tr} \, \mathbf{G}\left [ \mathbf{\eta},\mathbf{H}(s) \right ],
\label{derivative1}
\end{equation}
where I used Eq.~(\ref{genericU}) to replace the derivative.  
Using the cyclic property of traces,  $\mathrm{tr}\, \mathbf{A\, B} = \mathrm{tr}\, \mathbf{B \, A}$, 
one can rewrite the right-hand side of (\ref{derivative1}) 
\begin{eqnarray}
 \mathrm{tr} \, \mathbf{G}\left [ \mathbf{\eta},\mathbf{H}(s) \right ] 
 =  \mathrm{tr} \, \mathbf{G} \mathbf{\eta}\mathbf{H}(s)  - 
 \mathrm{tr} \, \mathbf{G} \mathbf{H}(s) \mathbf{\eta}  \nonumber \\
 = \mathrm{tr} \, \mathbf{\eta}\mathbf{H}(s)  \mathbf{G}  - 
 \mathrm{tr} \, \mathbf{\eta} \mathbf{G} \mathbf{H}(s) =\mathrm{tr}\, \mathbf{\eta} \left [\mathbf{H}(s), \mathbf{G}  \right ]
  \end{eqnarray}
which is maximized when the antisymmetric matrix
$\mathbf{\eta}=  \left [\mathbf{H}(s), \mathbf{G} \right ]^T = \left [\mathbf{G} ,\mathbf{H}(s) \right ]$. Standard SRG is defined by \textit{maximizing} the 
rate $\mathbf{H}(s)$ approaches the generator $\mathbf{G}$,  forcing $\mathbf{H}(s)$ to be \textit{as similar as possible} to $\mathbf{G}$.

This explains  the behavior of mixed symmetries under standard SRG: by forcing $\mathbf{H}$ to be 
as similar as possible to the SU(3) Casimir $-\mathbf{C}_2$, standard SRG tries to match extremal eigenpairs: low-lying states of $\mathbf{H}$ are 
driven to be like high-weight states of the Casimir.

\begin{figure}
\centering
\includegraphics[width=0.4\textwidth,clip]{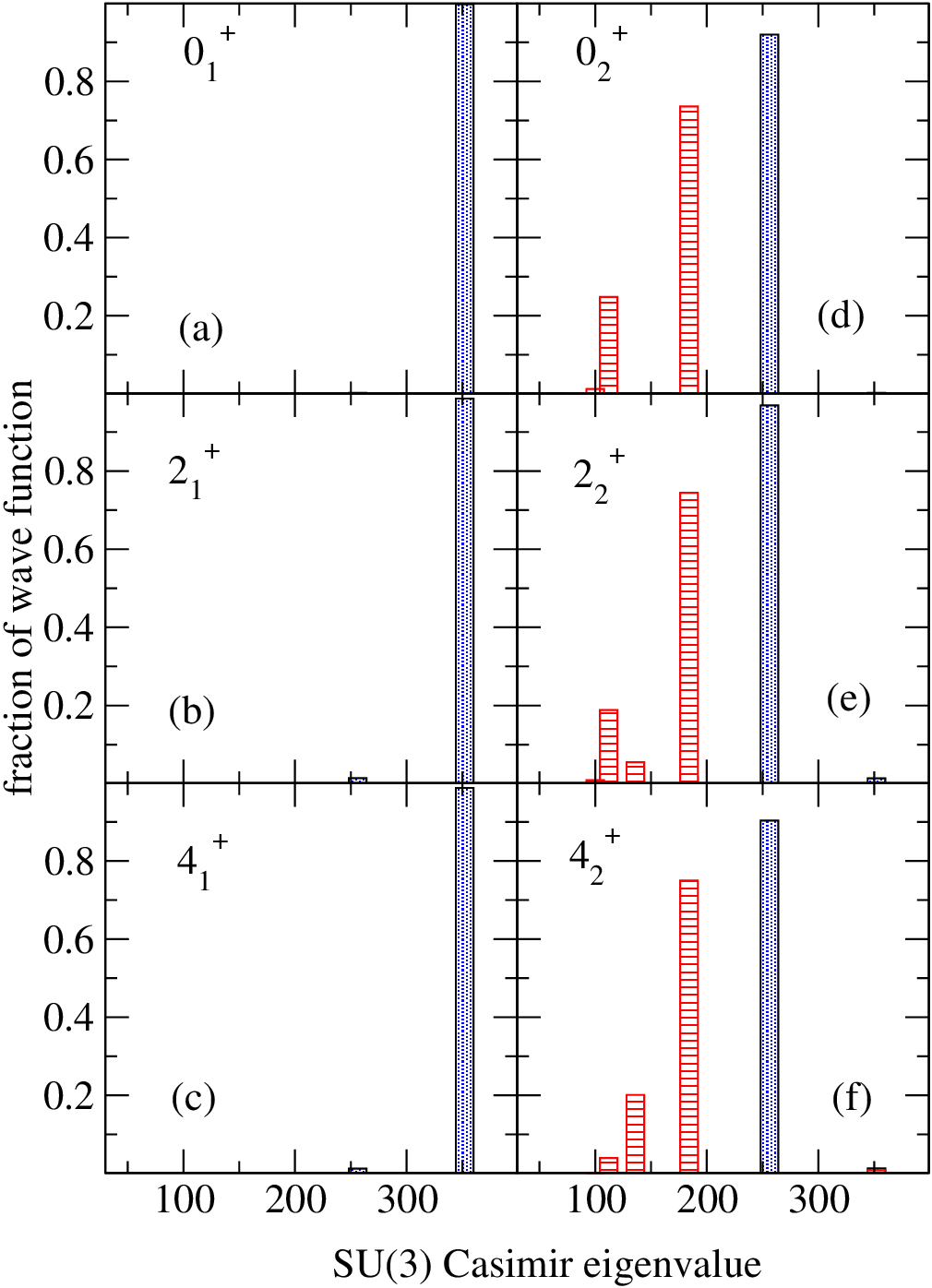}
\caption{ 
SU(3) decompositions for low-lying levels $^{36}$Ar in the $sd$ shell using 
the USDB interaction, evolved under two different SRG equations to $s=3$. Solid (blue) bars are for 
standard SRG, while (red) horizontal strips are for the new SRG.  For the ground state band, that is, 
the $0_1^+, 2_1^+, 4_1^+$ states, the two evolutions are indistinguishable.}
\label{ar36srgS3}
\end{figure}

I  now present an alternative. Recall that a dynamical symmetry is when a Hamiltonian merely \textit{commutes} with the Casmir(s)
of a group.  Thus, for my purposes here, a more appropriate condition is to maximize $\mathrm{tr} \, \left [ \mathbf{G}, \mathbf{H}(s) \right ]^\dagger 
\left [\mathbf{G}, \mathbf{H}(s)  \right]$, or, more practically, choose the evolution equation that maximizes its \textit{decrease}. Following the 
same methodology as before, one arrives at a modified SRG procedure, with 
\begin{equation}
\hat{\eta} = \left [ \left [ \left [ \mathbf{G}, \mathbf{H}(s) \right ], \mathbf{H}(s) \right ], \mathbf{G} \right  ] 
\label{newSRG}
\end{equation}
Properly coded, this only takes twice as much time as the original SRG.
The right column of Fig. ~\ref{ar36srg} shows the decomposition of the $2_2^+$ state under this `new' SRG.  Now the strength is pushed to irreps already in the plurality in the 
original decomposition. (The decomposition of the $0_1^+$ state under both SRGs is nearly indistinguishable.)

These results are not unique,  Fig.~\ref{ar36srgS3} shows the decomposition for the six lowest states, using both SRG equations, evolved to $s=3$. 
For the ground state band (left column, the $ 0_1^+, 2_1^+, 4_1^+$ states), the decompositions are indistinguishable and I show only the results from standard SRG. 
For the $ 0_2^+, 2_2^+, 4_2^+$ states, there is a difference, with standard SRG leading to the wave function being predominant in a higher-weight (and, compared to decomposition 
of states from the unevolved Hamiltonian, wrong) irrep, while the decomposition for states from the new SRG better reflect the unevolved state. Note that
under the new SRG a secondary component persists; evolving further to $s=5$, the results are little changed.  

\begin{figure}
\centering
\includegraphics[width=0.4\textwidth,clip]{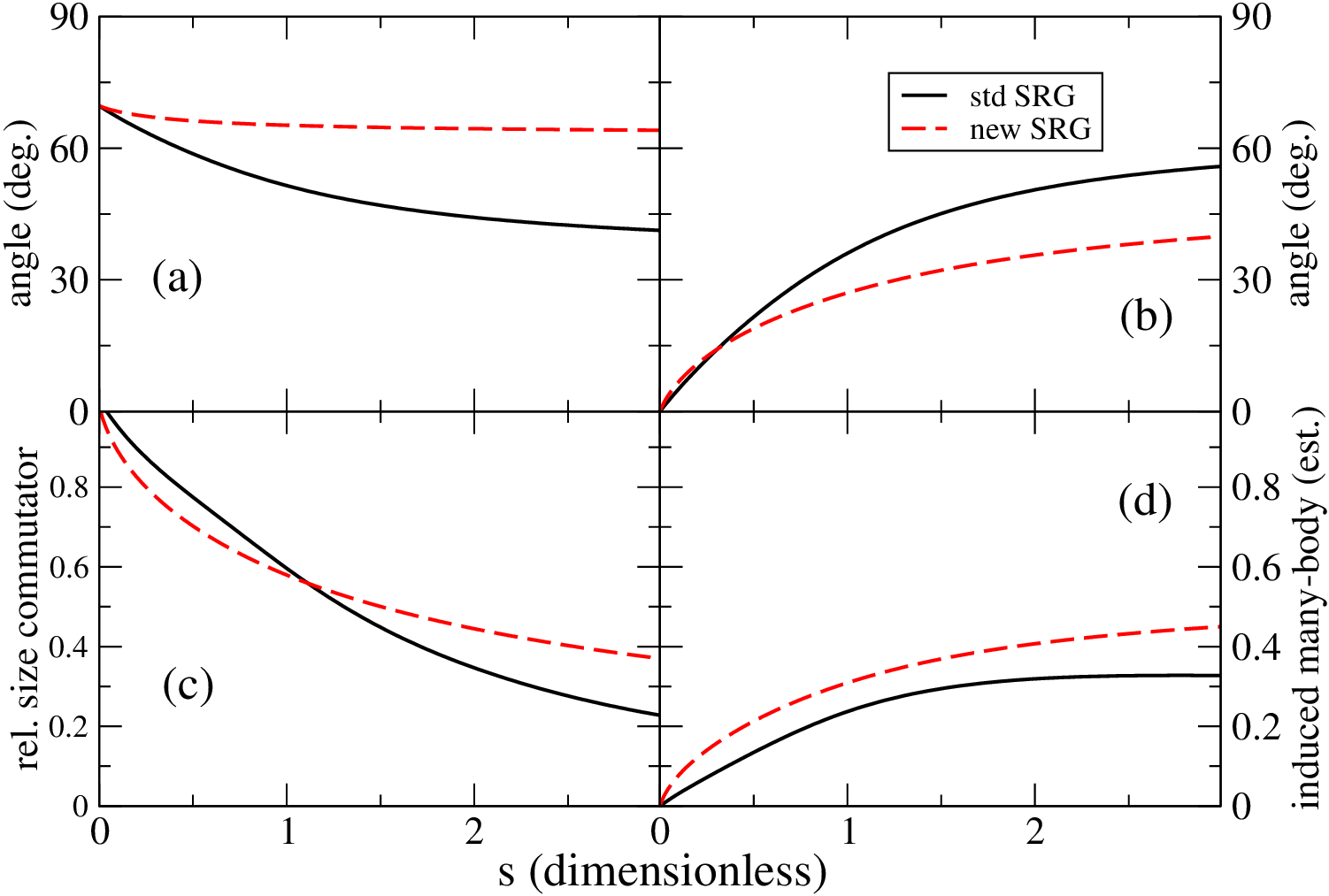}
\caption{ 
Tracking the evolution as a function of $s$ of the Hamiltonian for $^{36}$Ar, with solid (black) line for standard SRG and dashed (red) line for the new SRG.
(a) Angle, as defined by spectral distribution theory (see text for details)  between  evolved $\hat{H}(s)$ and the generator $\mathbf{G}$, here the quadratic SU(3) Casimir,
(b) Angle between evolved $\mathbf{H}(s)$ and original Hamiltonian $\mathbf{H}(0)$.  (c)  magnitude of the commutator $| [ \mathbf{H}(s) , \mathbf{G} ] |^2$, relative to magnitude at $s=0$. 
(d) Estimated fraction of $\hat{H}(s)$ with induced many-body terms. }
\label{evolution}
\end{figure}

More insight about the evolution can be gleaned from Fig.~\ref{evolution}.  Using the inner product (\ref{innerproduct}), one can calculate the angle between any two Hamiltonian-like 
operators.
Panel 4(a) shows the angle between the generator $\mathbf{G}$ (here -$\mathbf{C}_2$, the SU(3) Casimir) and the evolved Hamiltonian $\mathbf{H}(s)$, while panel 4(b) shows the angle between 
the evolved Hamiltonian and the original $\mathbf{H}(0)$, with solid black lines for the standard SRG and dashed (red) lines for the new SRG. In  both measurements, 
the new SRG evolves the Hamiltonian `less far away' than standard SRG.

This is confirmed in Table \ref{dot}, which gives the numerical overlap between the wave functions from the unevolved Hamiltonian, and Hamiltonians evolved by the standard and new SRG out to $s=3$. 
 It confirms that the ground state band, which is dominated by the highest weight irrep, has nearly identical evolution under both SRGs, but that for the $0^+_2, 2^+_2, 4^+_2$ states, 
 the new SRG leads to states with a much larger overlap with the unevolved states than  standard SRG. 

Because the motivation of the new SRG was to reduce the commutator $\left [ \mathbf{H}(s), \mathbf{G} \right  ]$, panel (c) shows the magnitude of the commutator, normalized to 1 at $s=0$. 
The magnitude is computed using the 2-norm, but because the commutator is an antisymmetric matrix, this is the same as (\ref{innerproduct}) up to a minus sign.  The commutator for the 
new SRG indeed drops more rapidly at first, although  for large $s$ the standard SRG overtakes it. 

It is well-known that SRG induces many-body forces even when starting from purely two-body interactions. In my evolution I worked directly with the many-body Hamiltonian. 
Nonetheless, I estimated the amount of induced many-body forces.  At $s=0$, most of the matrix elements of $\mathbf{H}$ are in fact zero, due to the two-body nature
of the Hamiltonian.  For $s > 0$, I measured what fraction of $\mathrm{tr} \, \mathbf{H}^2$ came from those matrix elements that were originally zero.  Shown in panel (d) of 
Fig.~\ref{evolution}, this at the very least gives a lower-limit on the induced many-body interactions.  Unsurprisingly given the triple commutator of Eq.~(\ref{newSRG}), 
the new SRG induces a larger fraction of many-body components, but still of comparable size to the standard SRG.  
In practical applications one will likely have to either include the induced three-body terms\cite{PhysRevLett.103.082501} or carry out some effective procedure such as in-medium SRG, where one normal-orders 
three-body and higher order terms with respect to a reference state\cite{tsukiyama2011medium,hergert2016medium}.  Because such procedures work well with standard SRG, there is no reason to think it will be different here.

 \begin{table}
 \caption{Overlaps between configuration-interaction wave functions, calculated using the unevolved ($s=0$) Hamiltonian and evolved to $s=3$ using the standard and new SRG.}
 \label{dot}
 \begin{tabular}{|c|l|l|l|l|l|l|}
 \hline
   & $0_1^+$ &  $2_1^+$ &  $4_1^+$ &$0_2^+$ &  $2_2^+$ &  $4_2^+$ \\
  $ \langle \psi(s=0) | \psi_\mathrm{std} (s=3) \rangle$ & 0.669 & 0.719  & 0.717 & 0.008  & 0.336 & 0.382 \\
   \hline
   $\langle \psi(s=0) | \psi_\mathrm{new} (s=3) \rangle$ &  0.643 & 0.696 & 0.702  & 0.561 & 0.695 & 0.836 \\
   \hline
   $\langle \psi_\mathrm{std}(s=3) | \psi_\mathrm{new} (s=3) \rangle$ & 0.999 & 0.992  & 0.991 &  0.007 & 0.201 &  0.170  \\
 
  \hline
 \end{tabular}
 \end{table}

Although I only show the case of $^{36}$Ar, other nuclides show similar behavior. 
When the decomposition is highly fragmented and the wave function from the unevolved Hamiltonian is not dominated by a single irrep, 
both SRG procedures will still drive the decomposition to a single irrep: the original SRG will still evolve to a high-weight irrep, even if that 
irrep had a small occupation originally, while under the new SRG the evolved dominant irrep tends to be near the average of the 
previously occupied irreps. The implications of this behavior will be investigated in future work. 

In summary, I have shown how to construct a unitary transformation that undoes mixed symmetries, by transforming away symmetry-mixing terms, leading to a system with nearly pure dynamical symmetry.    
Furthermore, I  introduced and demonstrated a new version of SRG that, at least in some aspects, provides superior behavior over standard SRG. 
One possible application beyond `unmixing' symmetries would be in \textit{symmetry-adapted} structure 
calculations \cite{0954-3899-35-12-123101}, which rely upon the wave functions being dominated by a few irreps; by reducing the fragmentation into other irreps, such calculations could be closer to the full-space results. 
Because this new SRG decouples differently from, and in some cases demonstrably better than standard SRG, it may have applications beyond unmixing symmetries.

This material is based upon work supported by the U.S. Department of Energy, Office of Science, Office of Nuclear Physics, 
under Award Number  DE-FG02-96ER40985.

\bibliographystyle{apsrev4-1}
\bibliography{johnsonmaster}

\end{document}